\documentclass{PoS}
\title{
$\rho$ meson decay from the lattice
}
\ShortTitle{
$\rho$ meson decay from the lattice
}
\author{
CP-PACS collaboration:
}
\author{
S.~Aoki,$^{a,b}$
M.~Fukugita,$^c$
K.-I.~Ishikawa,$^d$
\speaker{N.~Ishizuka},$^{a,e}$\thanks{E-mail : ishizuka@ccs.tsukuba.ac.jp}
K.~Kanaya,$^a$
Y.~Kuramashi,$^{a,e}$
Y.~Namekawa,$^{f}$
M.~Okawa,$^d$
K.~Sasaki,$^{e}$
A.~Ukawa,$^{a,e}$
T.~Yoshi\'e$^{a,e}$
\\
\\
\\
\llap{$^a$}
Graduate School of Pure and Applied Sciences,
University of Tsukuba, Tsukuba, Ibaraki 305-8571, Japan
\\
\llap{$^b$}
Riken BNL Research Center,
Brookhaven National Laboratory,
Upton, New York 11973, USA
\\
\llap{$^c$}
Institute for Cosmic Ray Research,
University of Tokyo,
Kashiwa 277-8582, Japan
\\
\llap{$^d$}
Department of Physics,
Hiroshima University,
Higashi-Hiroshima, Hiroshima 739-8526, Japan
\\
\llap{$^e$}
Center for Computational Sciences,
University of Tsukuba,
Tsukuba, Ibaraki 305-8577, Japan
\\
\llap{$^f$}
Department of Physics,
Nagoya University,
Nagoya 464-8602, Japan
}
\FullConference{
XXIVth International Symposium on Lattice Field Theory \\
July 23-28, 2006 \\
Tucson, Arizona, USA
}
\abstract{
We present preliminary results on the $\rho$ meson decay width
estimated from the scattering phase shift of the $I=1$ two-pion system.
The phase shift is calculated by the finite size formula
for non-zero total momentum frame (the moving frame)
derived by Rummukainen and Gottlieb,
using the $N_f=2$ improved Wilson fermion action
at $m_\pi/m_\rho=0.41$ and $L=2.53\ {\rm fm}$.
}
\begin{document}
%
%
\section{ Introduction }
Lattice study of the $\rho$ meson decay
is an important step for understanding
of the dynamical aspect of hadron reactions induced by
strong interactions.
There are already three studies~\cite{rhpipi_GMTW,rhpipi_LD,rhpipi_MM}.
The earlier two studies employed the quenched approximation
ignores the decay into two ghost pions.
The most recent work,
while using the $N_f=2$ dynamical configurations,
concentrated on the $\rho\to\pi\pi$ transition amplitude
rather than the full matrix
including the $\rho\to\rho$ and $\pi\pi\to\pi\pi$ amplitudes.
All three studies were carried out at an unphysical kinematics
$m_\pi/m_\rho > 1/2$.

In this work we attempt to carry out a more rigorous approach.
We estimate the decay width
from the scattering phase shift for the $I=1$ two-pion system.
The finite size formula
presented by Rummukainen and Gottlieb~\cite{fm_RG} is employed
for an estimation of the phase shift.
Calculations are carried out with $N_f=2$ full QCD configuration
previously generated for a study of light hadron spectrum
with a renormalization group improved gauge action
and a clover fermion action at $\beta=1.8$, $\kappa=0.14705$
on a $12^3\times 24$ lattice~\cite{conf_NMC}.
The parameters determined from the spectrum analysis
are $1/a=0.92\ {\rm GeV}$,
$m_\pi/m_\rho=0.41$, and $L=2.53\ {\rm fm}$.
All calculations of this work are carried out on
VPP5000/80 at the Academic Computing and Communications Center of
University of Tsukuba.
%
%
\section{ Method }
In order to realize a kinematics such that the energy of the two-pion state
is close to the resonance energy $m_\rho$,
we consider the non-zero total momentum frame (the moving frame)~\cite{fm_RG}
with the total momentum ${\bf p}=2\pi / L \cdot {\bf e}_3$.
The initial $\rho$ meson is assigned a polarization vector parallel to ${\bf p}$.
One of the final two pions carry the momentum ${\bf p}$,
while the other pion is at rest.
The energies ignoring hadron interactions are then given by
$E_1^0 = \sqrt{ m_{\pi }^2 + p^2 } + m_\pi$ for the two-pion state and
$E_2^0 = \sqrt{ m_{\rho}^2 + p^2 }        $ for the $\rho$ meson.
We neglect higher energy states
whose energies are much higher than $E_1^0$ and $E_2^0$.
On our full QCD configurations,
the invariant mass for the two-pion state
takes $\sqrt{s} = 0.97\times m_\rho$,
while $1.47\times m_\rho$
is expected for the zero total momentum.
The $\rho$ meson at zero momentum cannot decay energetically,
so that it can be used to extract $m_\rho$.

The hadron interactions shift the energy from $E_n^0$ to $E_n$ ($n=1,2$).
These energies $E_n$ are related to the two-pion scattering phase shift
$\delta(\sqrt{s})$ through the Rummukainen-Gottlieb formula~\cite{fm_RG},
which is an extension of the L\"uscher formula~\cite{fm_LU}
to the moving frame.
The formula for the total momentum ${\bf p}=p {\bf e}_3$
and the ${\bf A}_2^-$ representation of the rotation group
on the lattice reads
\begin{equation}
\frac{1}{\tan\delta(\sqrt{s})}
= \frac{ 1 }{ 2\pi^2 q \gamma }\
  \sum_{ {\bf r} \in \Gamma }  \
     \frac{ 1 + ( 3 r_3^2 - r^2 )/q^2 }{ r^2 - q^2 }
\ ,
\label{eq:RG_F}
\end{equation}
where $\sqrt{s} = \sqrt{ E^2 - p^2 }$
is the invariant mass,
$k$ is the scattering momentum
($\sqrt{s} = 2 \sqrt{ m_\pi^2 + k^2 }$),
$\gamma$ is the Lorentz boost factor ($\gamma=E/\sqrt{s}$),
and $q=k L / (2\pi)$.
The summation for ${\bf r}$ in (\ref{eq:RG_F})
runs over the set
\begin{equation}
  \Gamma
   = \{ {\bf r} |
        \ r_1 = n_1 \ ,
        \ r_2 = n_2 \ ,
        \ r_3 = ( n_3 + \frac{p}{2} \frac{L}{2\pi} )/\gamma \ ,
        \ {\bf n}\in {\bf Z}^3 \}
\ .
\end{equation}
The right hand side of (\ref{eq:RG_F})
can be evaluated by the method described in Ref.~\cite{phsh_YAMA}.

In order to calculate $E_1$ and $E_2$
we construct a $2\times 2$ matrix time correlation function,
\begin{equation}
G(t) =
\left(
\begin{array}{llll}
      \displaystyle \langle 0 | \ (\pi\pi)^\dagger(t) \ (\pi\pi) (t_s) \ | 0 \rangle
& & & \displaystyle \langle 0 | \ (\pi\pi)^\dagger(t) \   \rho_3 (t_s) \ | 0 \rangle  \\
      \displaystyle \langle 0 | \   \rho_3^\dagger(t) \ (\pi\pi) (t_s) \ | 0 \rangle
& & & \displaystyle \langle 0 | \   \rho_3^\dagger(t) \   \rho_3 (t_s) \ | 0 \rangle  \\
\end{array}
\right)
\ .
\label{eq:G}
\end{equation}
Here,
$\rho_3(t)$ is an interpolating operator
for the neutral $\rho$ meson with the momentum
${\bf p}=2\pi/L\cdot {\bf e}_3$
and the polarization vector parallel to ${\bf p}$;
$(\pi\pi)(t)$ is an interpolating operator for the two pions given
by
\begin{equation}
(\pi\pi)(t) = \frac{1}{\sqrt{2}}
    \Bigl(
         \pi^{-}({\bf p},t) \pi^{+}({\bf 0},t)
       - \pi^{+}({\bf p},t) \pi^{-}({\bf 0},t)
    \Bigl)
\ ,
\label{eq:pp_op}
\end{equation}
which belongs to the ${\bf A}_2^-$
and iso-spin representation with $I=1$, $I_z=0$.

We can extract the two energy eigenvalues
by a single exponential fitting of
the two eigenvalues $\lambda_1 (t,t_R)$ and $\lambda_2 (t,t_R)$
of the normalized matrix $M(t,t_R) = G(t) G^{-1}(t_R)$
with some reference time $t_R$~\cite{method_diag}
assuming that the lower two states dominate the correlation function.

In order to construct the meson state with non-zero momentum
we introduce a $U(1)$ noise $\xi_j({\bf x})$ in three-dimensional space
whose property is
\begin{equation}
\frac{1}{N_R}
\sum_{j=1}^{N_R}
\xi_j^\dagger ({\bf x}) \xi_j ({\bf y})
= \delta^3( {\bf x} - {\bf y} )
             \qquad \mbox{ for \ $N_R \to \infty$ }
\ .
\label{eq:xi_prop}
\end{equation}
We calculate the quark propagator
\begin{equation}
Q( {\bf x}, t | {\bf q}, t_s, \xi_j )
  = \sum_{\bf y} ( D^{-1} )({\bf x}, t ; {\bf y}, t_s )
    \cdot \Bigl[ \ {\rm e}^{ i{\bf q}\cdot{\bf y}} \xi_j({\bf y}) \ \Bigr]
\ ,
\label{eq:QP_Q}
\end{equation}
regarding the term in the square bracket as the source.
The two point function of the meson with the spin content $\Gamma$
and the momentum ${\bf p}$ can be constructed from $Q$ as
\begin{equation}
\frac{1}{N_R} \sum_{j=1}^{N_R} \
\sum_{\bf x} {\rm e}^{ - i {\bf p}\cdot {\bf x} }
\cdot
\Bigl\langle
  \ \gamma_5
  \ Q^\dagger ( {\bf x}, t | {\bf 0}, t_s, \xi_j ) \ \gamma_5 \Gamma^\dagger
  \ Q         ( {\bf x}, t | {\bf p}, t_s, \xi_j ) \          \Gamma
  \
\Bigr\rangle
\ ,
\end{equation}
where the bracket refers to the trace for color and spin indeces.

The quark contraction for the $\pi\pi\to\pi\pi$
and the $\pi\pi\to\rho$ components
of $G(t)$ are given by
\begin{equation}
\includegraphics[width=14.5cm]{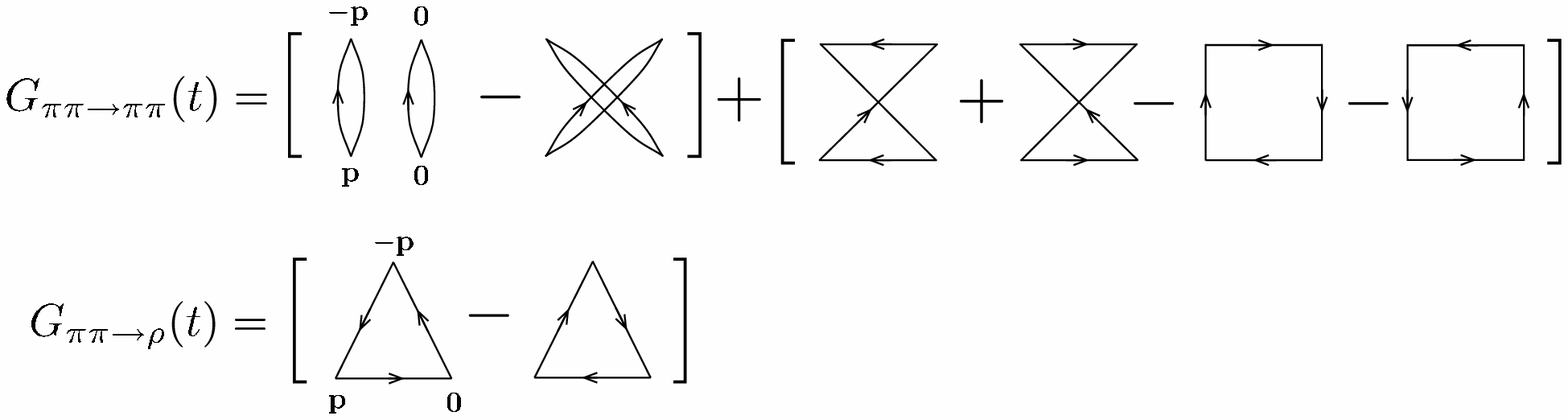}
\label{eq:quark_c}
\end{equation}
where
the four  verteces for the $\pi\pi\to\pi\pi$
and three verteces for the $\pi\pi\to\rho$ components
refer to the pion or the $\rho$ meson with definite momentum.
The time direction is upward in the diagrams,
and the $\rho\to\pi\pi$ component is given by changing the time direction.

The first term of the $\pi\pi\to\pi\pi$ component in (\ref{eq:quark_c})
can be calculated by introducing another $U(1)$ noise $\eta_j({\bf x})$
having the same property as $\xi_j({\bf x})$ in (\ref{eq:xi_prop});
\begin{equation}
\frac{1}{N_R}
\sum_{j=1}^{N_R}
\sum_{{\bf x}, {\bf y}} {\rm e}^{ - i {\bf p}\cdot {\bf x} }
    \cdot
    \Bigl\langle
         Q^\dagger ( {\bf x}, t | {\bf 0}, t_s, \xi_j  ) \
         Q         ( {\bf x}, t | {\bf p}, t_s, \xi_j  ) \Bigr\rangle
    \Bigl\langle
         Q^\dagger ( {\bf y}, t | {\bf 0}, t_s, \eta_j ) \
         Q         ( {\bf y}, t | {\bf 0}, t_s, \eta_j ) \Bigr\rangle
\ .
\label{eq:D1_quark_c}
\end{equation}
The second term of (\ref{eq:quark_c})
is obtained by exchanging the momentum of the sink in (\ref{eq:D1_quark_c}).

In order to construct the other terms of (\ref{eq:quark_c})
we calculate a quark propagator of another type by the source method,
\begin{equation}
W( {\bf x}, t | {\bf k}, t_1 | {\bf q}, t_s, \xi_j )
= \sum_{\bf z} ( D^{-1} )({\bf x}, t ; {\bf z}, t_1 )
    \cdot \Bigl[ {\rm e}^{ i{\bf k}\cdot{\bf z}}
                  \gamma_5 \ Q( {\bf z}, t_1 | {\bf q}, t_s, \xi_j )
          \Bigr]
\ ,
\label{eq:QP_W}
\end{equation}
where the term in the square bracket is regarded
as the source in solving the propagator.
Using $W$ we can construct the third to sixth terms
in the $\pi\pi\to\pi\pi$ component of (\ref{eq:quark_c})
by
\begin{eqnarray}
\mbox{ 3rd } &=&
\frac{1}{N_R}
\sum_{j=1}^{N_R}
\sum_{\bf x}  {\rm e}^{ - i {\bf p}\cdot {\bf x} } \cdot
    \Bigl\langle
      \  W^\dagger ( {\bf x}, t | {\bf 0}, t_s | - {\bf p}, t_s, \xi_j ) \
         W         ( {\bf x}, t | {\bf 0}, t   |   {\bf 0}, t_s, \xi_j ) \
    \Bigr\rangle
\ ,
\cr
\mbox{ 4th } &=&
\frac{1}{N_R}
\sum_{j=1}^{N_R}
\sum_{\bf x}  {\rm e}^{ - i {\bf p}\cdot {\bf x} } \cdot
    \Bigl\langle
      \  W         ( {\bf x}, t | {\bf 0}, t_s |   {\bf p}, t_s, \xi_j ) \
         W^\dagger ( {\bf x}, t | {\bf 0}, t   |   {\bf 0}, t_s, \xi_j ) \
    \Bigr\rangle
\ ,
\cr
\mbox{ 5th } &=&
\frac{1}{N_R}
\sum_{j=1}^{N_R}
\sum_{\bf x}  {\rm e}^{ - i {\bf p}\cdot {\bf x} } \cdot
    \Bigl\langle \
      \  W         ( {\bf x}, t | {\bf p}, t_s | {\bf 0}, t_s, \xi_j ) \
         W^\dagger ( {\bf x}, t | {\bf 0}, t   | {\bf 0}, t_s, \xi_j ) \
    \Bigr\rangle
\ ,
\cr
\mbox{ 6th } &=&
\frac{1}{N_R}
\sum_{j=1}^{N_R}
\sum_{\bf x}  {\rm e}^{ - i {\bf p}\cdot {\bf x} } \cdot
    \Bigl\langle \
      \  W^\dagger ( {\bf x}, t | - {\bf p}, t_s | {\bf 0}, t_s, \xi_j ) \
         W         ( {\bf x}, t |   {\bf 0}, t   | {\bf 0}, t_s, \xi_j ) \
    \Bigr\rangle
\ .
\label{eq:QC_B}
\end{eqnarray}
The two terms of $\pi\pi\to\rho$ of (\ref{eq:quark_c})
can be similarly constructed by
\begin{eqnarray}
\mbox{ 1st } &=&
\frac{1}{N_R}
\sum_{j=1}^{N_R}
\sum_{\bf x}  {\rm e}^{ - i {\bf p}\cdot {\bf x} } \cdot
  \Bigl\langle
    \  W^\dagger ( {\bf x}, t | - {\bf p}, t_s | {\bf 0}, t_s, \xi_j ) \ (\gamma_5 \gamma_3 ) \
       Q         ( {\bf x}, t                  | {\bf 0}, t_s, \xi_j ) \
  \Bigr\rangle
\ ,
\cr
\mbox{ 2nd } &=&
\frac{1}{N_R}
\sum_{j=1}^{N_R}
\sum_{\bf x}  {\rm e}^{ - i {\bf p}\cdot {\bf x} } \cdot
  \Bigl\langle
    \  Q^\dagger ( {\bf x}, t                | {\bf 0}, t_s, \xi_j ) \ (\gamma_5 \gamma_3 ) \
       W         ( {\bf x}, t | {\bf p}, t_s | {\bf 0}, t_s, \xi_j ) \
  \Bigr\rangle
\ .
\label{eq:QC_pprh}
\end{eqnarray}

In this work we set the source at $t_s=4$
and impose the Dirichlet boundary condition in the time direction.
We calculate the $Q$-type propagators
for four sets of ${\bf q}$ and the $U(1)$ noise in (\ref{eq:QP_Q})~:
$
( {\bf q}, {\rm noise})=\{
( {\bf 0}, \xi  ) ,
( {\bf 0}, \eta ) ,
( {\bf p}, \xi  ) ,
(-{\bf p}, \xi  )       \}$.
The $W$-type propagators are calculated
for 22 sets of
${\bf k}$, $t_1$ and ${\bf q}$ in (\ref{eq:QP_W})~:
$
(  {\bf k}, t_1      |  {\bf q} )=\{
(  {\bf p}, t_s      |  {\bf 0} ) ,
( -{\bf p}, t_s      |  {\bf 0} ) ,
(  {\bf 0}, t_s      |  {\bf p} ) ,
(  {\bf 0}, t_s      | -{\bf p} ) ,
(  {\bf 0}, t_1=4-21 |  {\bf 0} ) \}$,
with the same $U(1)$ noise $\xi$.
All diagrams for the time correlation function
can be calculated with combinations of these propagators.
We choose $N_R=10$ for the number of $U(1)$ noise.
We carry out additional measurements to reduce statistical errors
using the source operator is located at $t_s+T/2$
and the Dirichlet boundary condition is imposed at $T/2$.
We average over the two measurements for the analysis.
Thus we calculate $520$ quark propagators for each configuration.
The total number of configurations analyzed are 800 separated by 5
trajectories~\cite{conf_NMC}.
%
%
\section{ Results }
\begin{figure}[t]
\begin{center}
\includegraphics[width=7.5cm]{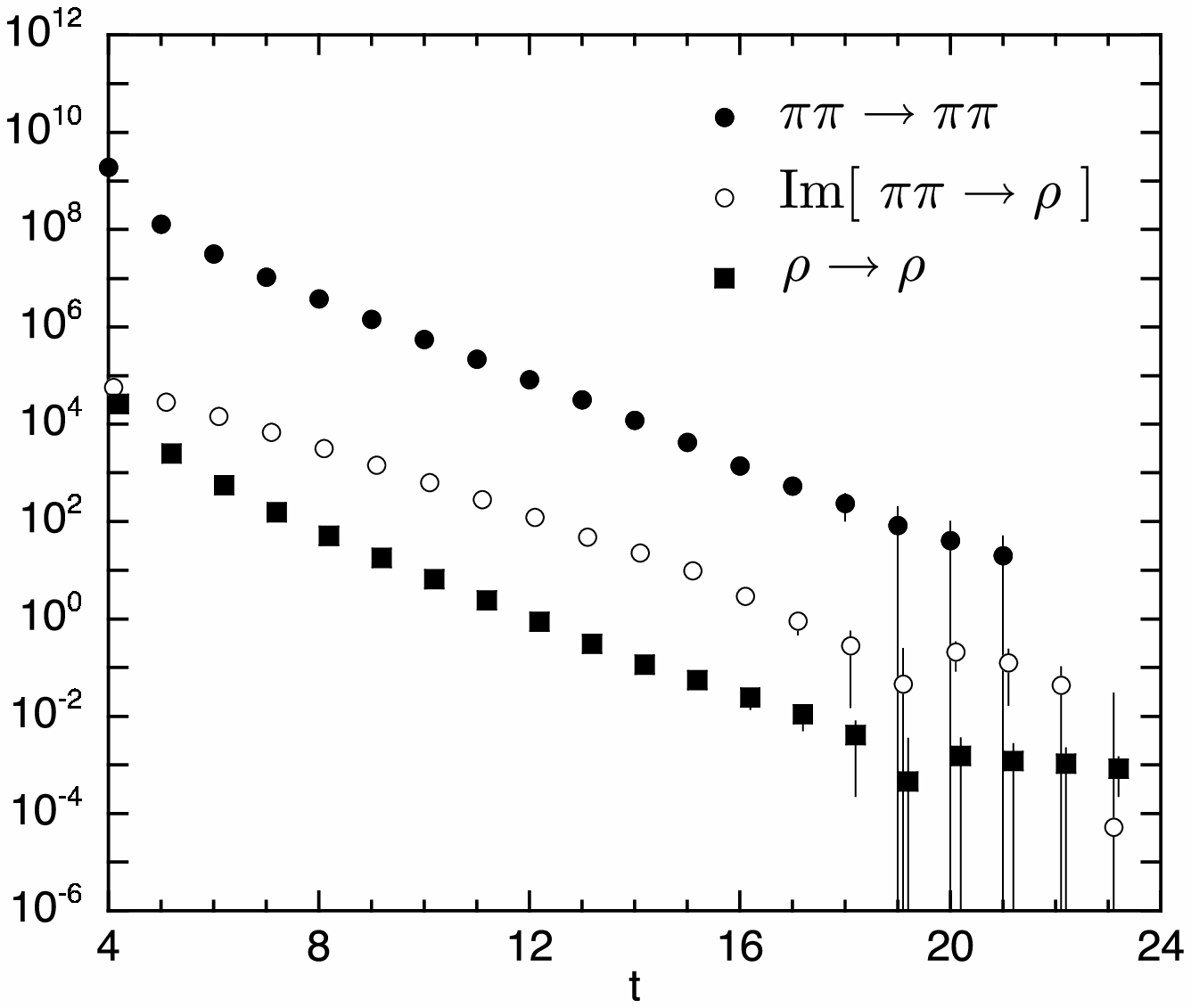}
\includegraphics[width=7.5cm]{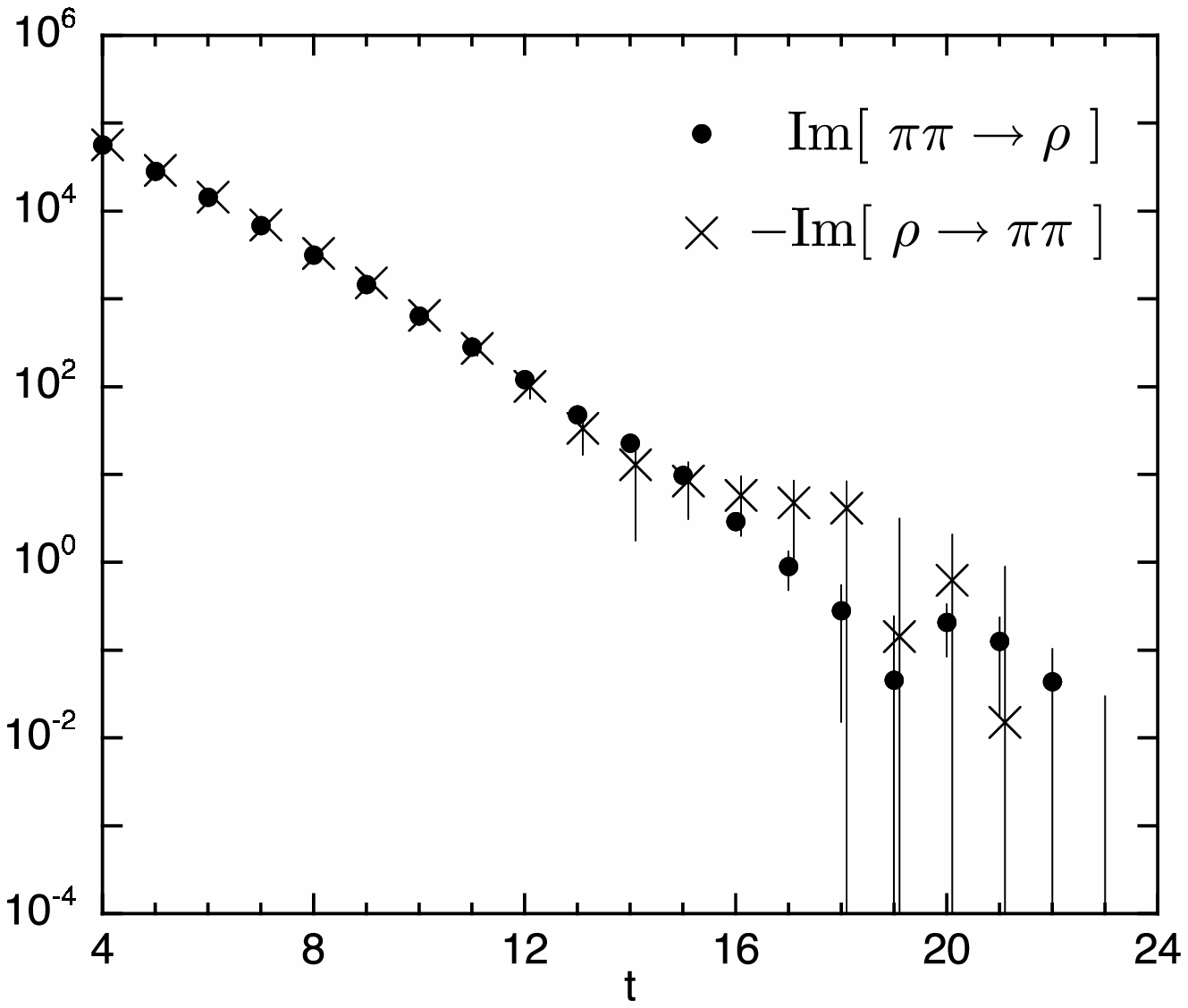}
\end{center}
\vspace{-0.8cm}
\caption{
$G(t)$
\label{fig:G_t}
}
\end{figure}

In Fig.~\ref{fig:G_t}
we plot the real part of the diagonal components ($\pi\pi\to\pi\pi$ and $\rho\to\rho$)
and the imaginary part of the off-diagonal components ($\pi\pi\to\rho$, $\rho\to\pi\pi$)
of $G(t)$.
Our construction of $G(t)$ is such that
the sink and source operators are identical
for a sufficiently large number of the $U(1)$ noise.
In this case
we can prove that $G(t)$ is an Hermitian matrix
and the off-diagonal parts are pure imaginary from $P$ and $CP$ symmetry.
We find that this is valid within statistics,
but the statistical errors
of the $\rho\to\pi\pi$ component is larger than those of $\pi\pi\to\rho$
in Fig.~\ref{fig:G_t}.
In the following analysis
we substitute $\rho\to\pi\pi$ by $\pi\pi\to\rho$
to reduce the statistical error.
%

\begin{figure}[t]
\begin{center}
\includegraphics[width=7.5cm]{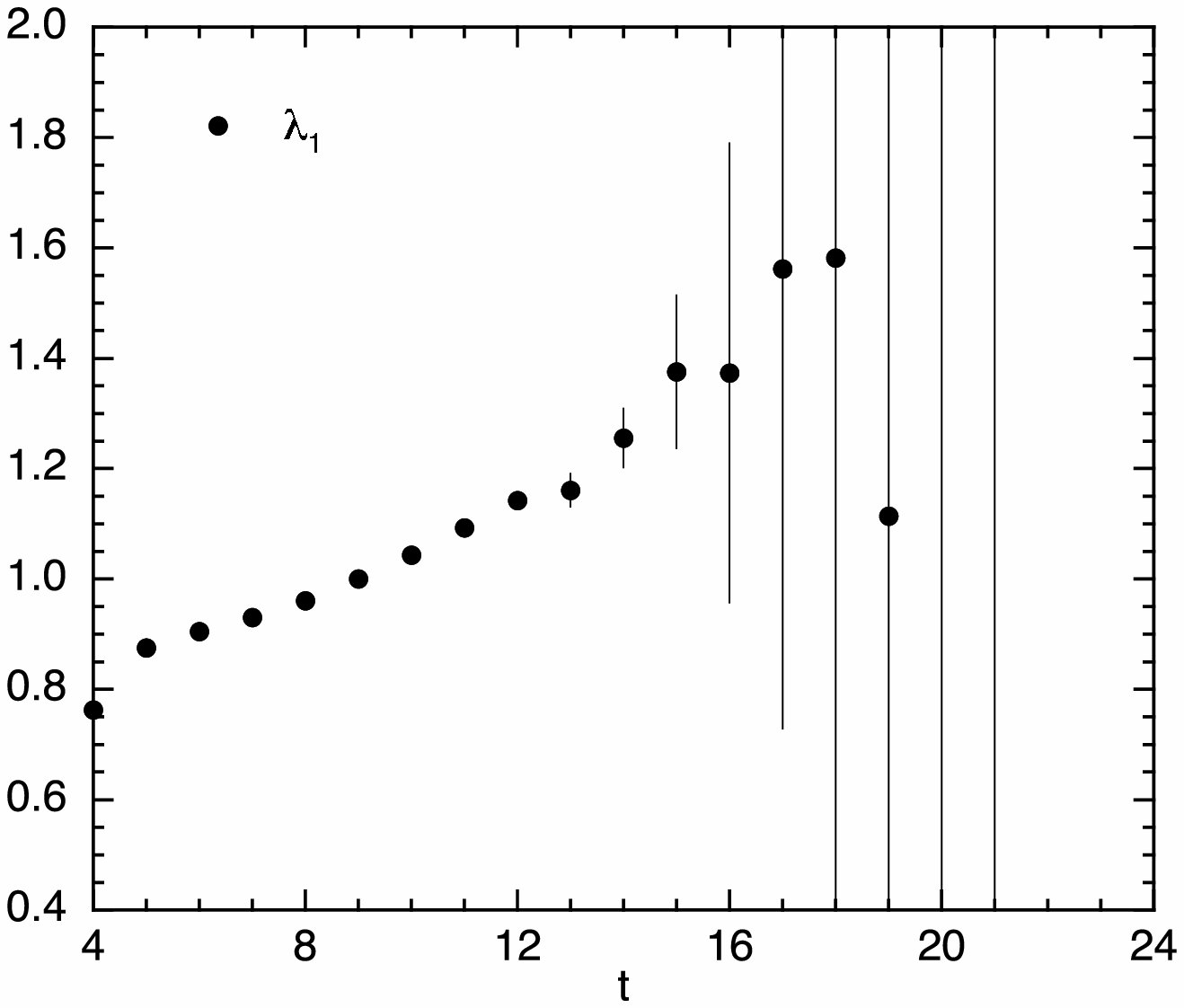}
\includegraphics[width=7.5cm]{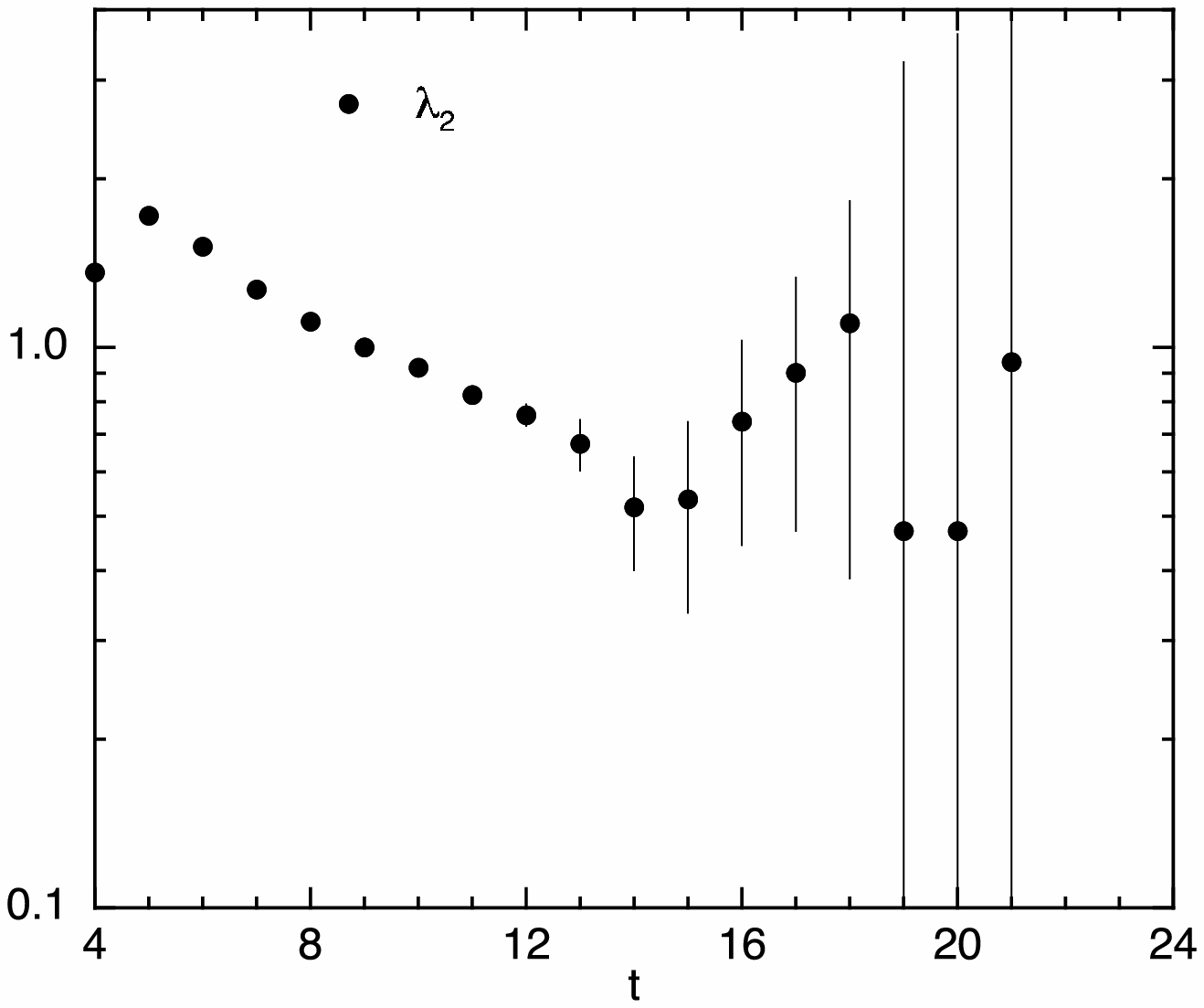}
\end{center}
\vspace{-0.8cm}
\caption{
Normalized eigenvalues $\lambda_1(t,t_R)$ and $\lambda_2(t,t_R)$.
\label{fig:Lambda_t}
}
\end{figure}

The two eigenvalues $\lambda_1(t,t_R)$ and $\lambda_2(t,t_R)$
for the matrix $M(t,t_R) = G(t) G^{-1}(t_R)$
are shown in Fig.~\ref{fig:Lambda_t}.
We set the reference time $t_R=9$ and
normalize the eigenvalues
by the correlation function for the free two-pion system,
$\langle 0| \pi(-{\bf p},t) \pi({\bf p},t_s) |0\rangle
 \langle 0| \pi( {\bf 0},t) \pi({\bf 0},t_s) |0\rangle
$.
Thus the slope of the figure corresponds to the energy difference
$\Delta E_n=E_n-E_1^0$.
We observe that the energy difference for $\lambda_1$ is negative and
that for $\lambda_2$ is positive.
This means that
the two-pion scatting phase shift
is positive for the lowest state and negative for the next higher state.
%

\begin{figure}[t]
\begin{center}
\includegraphics[width=9cm]{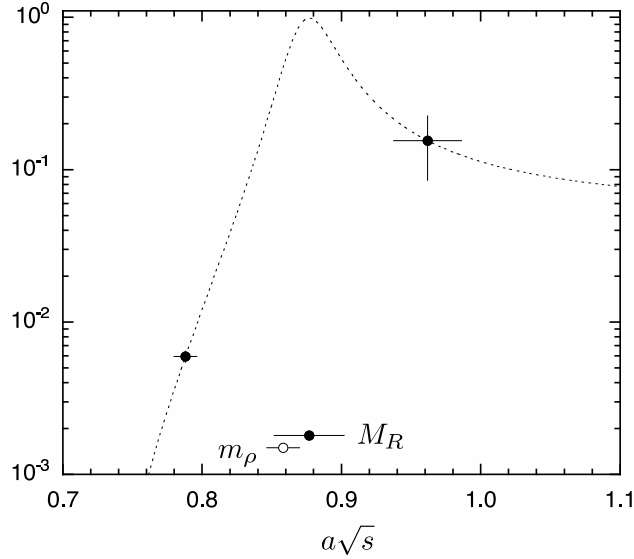}
\end{center}
\vspace{-0.8cm}
\caption{
$\sin^2\delta(\sqrt{s})$,
position of $m_\rho$ and resonance mass $M_R$.
\label{fig:sin2Del}
}
\end{figure}

We extract the energy difference $\Delta E_n$ for both states
by a single exponential fitting of the normalized eigenvalues
$\lambda_1$ and $\lambda_2$ for the time range $t=10-16$.
Then we reconstruct the energy $E_n$ in the moving frame
by adding the energy of the two free pions,
{\it i.e.,}
$E_n=\Delta E_n + E_1^0$,
and convert it to the invariant mass $\sqrt{s}$.
Substituting $\sqrt{s}$ into the Rummukainen-Gottlieb formula (\ref{eq:RG_F})
we obtain the scattering phase shift :
%
%
\hfill\break
\begin{equation}
\begin{array}{clccccc}
\hline
\hline
& \displaystyle a \sqrt{s}        &&&& \displaystyle  \tan\delta(\sqrt{s}) & \\
\hline
& \displaystyle 0.7880 \pm 0.0082 &&&& \displaystyle   0.0773 \pm 0.0033   & \\
& \displaystyle 0.962  \pm 0.024  &&&& \displaystyle  -0.43   \pm 0.12     & \\
\hline
\hline
\end{array}
\label{eq:tanD}
\end{equation}
\hfill\break
The $\rho$ meson mass obtained at zero momentum is
$a m_\rho =0.858 \pm 0.012$.
Hence
the sign of the scattering phase shifts at $\sqrt{s}< m_\rho$
is positive (attractive interaction)
and that at $\sqrt{s}> m_\rho$
is negative (repulsive interaction) as expected.
The corresponding results for $\sin^2\delta(\sqrt{s})$,
which is proportional to the scattering cross section of the two-pion system,
are plotted in Fig.~\ref{fig:sin2Del}
together with the position of $m_\rho$.
%

In order to estimate the $\rho$ meson decay width at the physical quark mass
we parameterize the scattering phase shift
by the effective $\rho\to\pi\pi$ coupling constant $g_{\rho\pi\pi}$,
\begin{equation}
\tan\delta(\sqrt{s})
 = \frac{ g_{\rho\pi\pi}^2 }{ 6\pi } \cdot
   \frac{ k^3 }{ \sqrt{s} ( M_R^2 - s ) }
\ ,
\label{eq:tanD_g}
\end{equation}
with $g_{\rho\pi\pi}$ defined by the effective Lagrangian,
\begin{equation}
L_{\rm eff.} = g_{\rho\pi\pi}
  \cdot \epsilon_{abc}
  ( k_1 - k_2 )_\mu   \rho_\mu^a(p) \pi^b(k_1) \pi^c(k_2)
\ ,
\end{equation}
where
$k$ is the scattering momentum and $M_R$ is the resonance mass.
The coupling $g_{\rho\pi\pi}$
generally depends on the quark mass and the energy, but our present data
at a single quark mass do not provide this information.
Here we assume that these dependences are small
and try to estimate $g_{\rho\pi\pi}$ and $M_R$
from our results in (\ref{eq:tanD}).
We also estimate the $\rho$ meson decay width at the physical quark mass
from
\begin{equation}
\Gamma_\rho
  = \frac{ g_{\rho\pi\pi}^2 }{ 6\pi } \cdot
    \frac{ \bar{k}_\rho^3 }{ \bar{m}_\rho^2 }
  = g_{\rho\pi\pi}^2 \times 4.128 \ \ {\rm MeV}
\ ,
\end{equation}
where
$\bar{m}_\rho$ is the $\rho$ meson mass at the physical quark mass
and $\bar{k}_\rho$ is the scattering momentum at $\sqrt{s}=\bar{m}_\rho$.

Our final results are as follows.
\begin{eqnarray}
a M_R          &=& 0.877 \pm 0.025           \cr
g_{\rho\pi\pi} &=& 6.01  \pm 0.63            \cr
\Gamma_\rho    &=& 149   \pm 31   \ \ {\rm MeV}
\ .
\label{eq:FinalR}
\end{eqnarray}
The resonance mass $M_R$ obtained from the scattering phase shift
is consistent with $a m_\rho=0.858\pm 0.012$
obtained from the $\rho$ meson with zero momentum.
The $\rho$ meson decay width $\Gamma_\rho$ at the physical quark mass
is consistent with experiment ($150\ {\rm MeV}$).
In Fig.~\ref{fig:sin2Del}
we indicate the position of $M_R$
and draw the line given by (\ref{eq:tanD_g})
with $g_{\rho\pi\pi}$ and $M_R$ in (\ref{eq:FinalR}).
%
%
\section{ Summary }
We have shown
that a direct calculation of the $\rho$ meson decay width
from the scattering phase shift for the $I=1$ two-pion system
is possible with present computing resources.
However, several issues remain which should be investigated
in future work.
The most important issue is a proper evaluation
of the quark mass and energy dependence of
the effective $\rho\to\pi\pi$ coupling constant $g_{\rho\pi\pi}$.
This constant is used to obtain the physical decay width
at $m_\pi/m_\rho=0.18$
from our results at $m_\pi/m_\rho=0.41$
by a long chiral extrapolation.
In principle
we can estimate the decay width from the scattering phase shift
without such a parameterization,
if we have data for several energy values
at or near the physical quark mass.
This will be our goal toward
the lattice determination of the $\rho$ meson decay.

This work is supported in part by Grants-in-Aid of the Ministry of Education
(Nos.
17340066, 
16540228, 
18104005, 
17540259, 
13135216, 
18540250, 
15540251, 
13135204, 
16740147, 
18740139  
).
The numerical calculations have been carried out
on VPP5000/80
at Academic Computing and Communications Center of University of Tsukuba.
%
%

%
%
\end{document}